\documentclass[10pt]{article} 

\usepackage{MnSymbol}  
\usepackage{graphicx}
\usepackage[margin=1in]{geometry}
\usepackage{multicol}

\newcommand{\av}[1]{\left<#1\right>}
\newcommand{\avav}[1]{\llangle #1 \rrangle} 
\newcommand{\zmax}{\ensuremath{{z_\mathrm{max}}}}

\begin{document}
  
\title{Thermodynamics of Surface-Bounded Exospheres}
\author{Norbert Sch\"orghofer, Planetary Science Institute, Honolulu, Hawaii, USA}
\maketitle

\begin{abstract}
  Neutral exospheres of the Moon, Mercury, and several other solar system bodies consist of particles on ballistic trajectories.
  Here, the vertical density profile of a surface-bounded exosphere is calculated using thermodynamic averages of an ensemble of ballistic trajectories.  
 When the initial velocities follow the so-called ``Maxwell-Boltzmann Flux" (MBF) distribution, the classical density profile results. For many other probability distributions, the density approaches infinity near the surface and has a more ground-hugging vertical profile than the classical solution.  Even MBF on a rough surface results in a ground-hugging component. Observed vertical density profiles that were interpreted as a superposition of a hot and a cold (ground hugging) population may in fact be consistent with a population at a single temperature.  
\end{abstract}

\begin{multicols}{2}

\section{Introduction}

The Moon, Mercury, and several other solar system bodies have rarefied atmospheres, which are collisionless exospheres (e.g., H, H$_2$, He, Na, Ar, and H$_2$O) where neutral atoms and molecules follow ballistic trajectories \cite{killen99rev,stern99rev}.
Surface-bounded exospheres are relevant to recent and upcoming measurement campaigns, such as by the LADEE (Lunar Atmosphere Dust and Environment Explorer) spacecraft and the MESSENGER and BepiColombo missions to planet Mercury.

Depending on the chemical species and the production mechanism, a particle is initially ejected by a thermal or non-thermal process. It then either immediately escapes to space or falls back onto the surface.
The vibrational frequency of the bond with the substrate surface is typically $10^{13}$~Hz \cite{dejong90}, so a particle should quickly thermalize, take on the temperature of the surface, leave with a thermal velocity distribution, and undergo a sequence of ballistic hops \cite{watson61b,hurley16}.
Some, usually minor, fraction will bounce elastically instead \cite{haynes92}.
In any case, neutral exospheres consist of ballistic trajectories with an ensemble of launch velocities, and the vertical density profile of a steady-state exosphere is associated with a temperature.  
Temperature is defined in terms of the change in the number of accessible quantum states with energy \cite{kittelkroemer}; collisions among particles are not required to define temperature. The temperature of a gravitationally-bound exosphere is the temperature of the surface it is in equilibrium with.

Exospheres above a dense atmosphere, such as on Earth and the Sun, have long been investigated theoretically \cite{opik59,johnson60,aamodt62,chamberlain63,shen63}.
For the simple case of constant gravitational acceleration $g$ and a scale height much smaller than the radius of the body, the density follows an exponential dependence 
\begin{equation}
\rho(z) = \rho(0) e^{-z/H}
\label{baro}
\end{equation}
where $z$ is the height above the exobase and $H$ the scale height:
\begin{equation}
  H =  \frac{kT}{mg}
  \label{H}
\end{equation}
where $k$ is the Boltzmann constant, $T$ the temperature associated with the initial velocities, and $m$ the mass of the atom or molecule. This is the same scale height as that of an isothermal hydrostatic atmosphere, and eq.\ (\ref{baro}) is the ``barometric law'', and in the context of an exosphere often referred to as (the simplest form of) the ``Chamberlain distribution''.  To re-iterate, hydrostatic equilibrium is not assumed in the derivation of this solution.

Based on numerical simulations, it was soon realized that launch velocities with a Maxwell-Boltzmann (MB) distribution do not result in a barometric exosphere \cite{smith78,hodges80b}, and it was proposed that the launch velocities for an exobase correspond to a ``Maxwell-Boltzmann flux'' (MBF) distribution, which places an extra factor of the vertical velocity component, $v_z$, in front of the MB distribution. As shown below analytically, this indeed results in an exponential density profile.

When the base of the exosphere is a solid surface, it is not clear whether using the MBF distribution is still justified, and various velocity distributions can result from different production or desorption processes.
Moreover, a surface-bounded exosphere involves a fixed or limited number of particles whereas an exobase functions as a reservoir of particles, so the population in the exosphere is not a closed system.
Here, a first-principle approach is used that is ideally suited for the surface-bounded case and fully analytically tractable.

\section{Thermodynamic averages of ballistic trajectories}

\subsection{A single ballistic hop}
Let $v_z$ denote the initial vertical velocity component.
It follows from elementary mechanics that for constant $g$ the duration of ballistic flight is
\begin{equation}
t_D = \frac{2 v_z}{g}
\label{tD}
\end{equation}
and the maximum height of a ballistic trajectory is
\begin{equation}
\zmax = \frac{v_z^2}{2g}
\label{zmax}
\end{equation}
The vertical velocity as a function of time and height, respectively, is
\begin{equation}
\frac{dz}{dt} = v_z - gt = \sqrt{v_z^2 - 2gz}
\end{equation}
The time the particle spends at a particular height $z$ is proportional to $1/|dz/dt|$.
Therefore, the density profile for a single ballistic hop is
\begin{equation}
  \rho(z) =  \frac{g}{v_z} \frac{1}{\sqrt{v_z^2 - 2gz}}
  \label{rhoof1}
\end{equation}
where the prefactor was determined from normalization,
\begin{equation}
\int_0^\zmax \rho(z) \, dz = 1
\end{equation}

\subsection{Ensemble averages}

Although an exosphere is collisionless, the trajectories still represent a thermodynamic ensemble (i.e., a statistical ensemble in equilibrium), as the particles thermalize with the surface.  
Given a probability distribution of initial velocities $P(v_z)$, the ensemble average of a quantity $X$ per hop is
\begin{equation}
\av{X} = \int X P(v_z) \, dv_z
\label{av}
\end{equation}
The time average for a stationary exosphere, $\avav{.}$, has to be weighted by the flight duration and is
\begin{equation}
  \avav{X} = \int X P(v_z) \frac{t_D}{\av{t_D}} \, dv_z 
  = \int X P(v_z) \frac{v_z}{\av{v_z}} \, dv_z
\label{avav}
\end{equation}
Both types of averages are properly normalized: $\av{1}=1$ and $\avav 1 = 1$.

Equations (\ref{rhoof1}), (\ref{av}), and (\ref{avav}) imply
  \begin{eqnarray}
  \av{\rho(0)} &=& g \int_0^\infty \frac{1}{v_z^2} P(v_z) \, dz \label{avrho0} \\
  \avav{\rho(0)} &=& \frac{g}{\av{v_z}} \int_0^\infty \frac{1}{v_z} P(v_z) \, dz  \label{avavrho0}
\end{eqnarray}
When the probability distribution is expanded for small $v_z$ as $P(v_z) = P(0) + P'(0) v_z + ...$, then $\avav{\rho(0)} < \infty$ requires $P(0)=0$. If $P(0)>0$, the surface density diverges logarithmically. For the vertical component of a Maxwell-Boltzmann distribution, $P(v_z=0)$ does not vanish, so the surface density diverges.
When $P(v_z)$ is continuous but not analytic in the neighborhood of zero, then $P(0)=0$ is a necessary, but not a sufficient condition for convergence.

To form the average density profile, the integration is over all velocities that are sufficiently high to reach a given height, i.e., $v_z > \sqrt{2gz}$:
\begin{equation}
  \av{\rho} = \int_{\sqrt{2gz}}^\infty \rho(z;v_z) P_M (v_z) \, dv_z 
  \label{defavrho}
\end{equation}
The time-averaged profile $\avav{\rho}$ is proportional to $\avav{1/|dz/dt|}$,
\begin{eqnarray}
  \avav{\rho} &=& \int \rho(z;v_z) P(v_z) \frac{v_z}{\av{v_z}} \, dv_z \\
  &=& \frac{g}{\av{v_z}} \int {1\over |dz/dt|} P(v_z)  \, dv_z \label{halfform}\\
  &=& \frac{g}{\av{v_z}} \int_{\sqrt{2gz}}^\infty {1\over \sqrt{v_z^2-2gz}} P(v_z)  \, dv_z
  \label{newform}
\end{eqnarray}

\section{Exact solutions}

\subsection{Maxwell-Boltzmann (MB) distribution: a divergent density profile}

The 3-dimensional MB distribution is the product of three Boltzmann distributions:
\begin{equation}
P_{3M} = 
\left( \sqrt{\beta\over\pi} e^{-\beta v_x^2} \right)
\left( \sqrt{\beta\over\pi} e^{-\beta v_y^2} \right)
\left( 2 \sqrt{\beta\over\pi}e^{-\beta v_z^2} \right)
\label{M_cart}
\end{equation}
with $\beta = m/(2kT)$.
The probability distribution of the vertical velocity component $v_z$ is
\begin{equation}
  P_M(v_z) = 2 \sqrt {s \over\pi} e^{-\beta v_z^2}
  \label{PM1}
\end{equation}
which has averages
\begin{eqnarray}
  \av{v_z}_M &=& \frac{1}{\sqrt{\pi \beta}}, \quad
  \av{v_z^2}_M = \frac{1}{2\beta} = \frac{k T}{m}  \\
  \av{v_z^3}_M &=& \frac{1}{ \sqrt{\pi \beta^3} } \\
  \avav{v_z}_M &=& \frac{1}{2}\sqrt{\pi\over \beta}, \quad
  \avav{v_z^2}_M = \frac{1}{\beta}
\end{eqnarray}

The averages of $t_D$ and \zmax, using eqs.\ (\ref{tD}), (\ref{zmax}), (\ref{av}), and (\ref{avav}) are
\begin{eqnarray}
  \av{t_D}_M &=& \frac{2}{g} \av{v_z}_M = \sqrt{8H \over g\pi}   \label{avtD} \\
  \av\zmax_M &=& \frac{\av{v_z^2}_M}{2g} = \frac{kT}{2mg} = \frac{H}{2} \label{avzmax1} \\
  \avav{t_D}_M &=& \frac{2}{g} \avav{v_z}_M = \sqrt{2H\pi \over g} 
  = \frac{\pi}{2} \av{t_D}_M \label{avavtD} \\
 \avav{\zmax}_M &=&  \frac{1}{g^2 \av{t_D}_M} \av{v_z^3}_M = H \label{avavzmax}
\end{eqnarray}
The average duration of a hop is given by eq.\ (\ref{avtD}), whereas the average flight duration of all particles in-flight at a given time is given by eq.\ (\ref{avavtD}). The average maximum height per hop is $H/2$ (\ref{avzmax1}), whereas the maximum height reached by the particles in flight at any given time is $H$ (\ref{avavzmax}).

To average of the density profile is
\begin{eqnarray}
  \av{\rho}_M &=& \int_{\sqrt{2gz}}^\infty \rho(z;v_z) P_M (v_z) \, dv_z  \\
  &=& \sqrt{\pi \over 4 z H} \,\mathrm{Erfc}\left(\sqrt{z\over H}\right)
  \label{avrho_M}
\end{eqnarray}
where Erfc is the complementary Error function.
This result has the correct normalization and average:
\begin{eqnarray}
\int_0^\infty \av\rho_M dz &=& 1 \\
\int_0^\infty z \av{\rho}_M dz &=& \frac H 3 \label{avz}
\end{eqnarray}
The median height is determined numerically as $z_m \approx 0.12 H$.

The time-averaged density profile is
\begin{eqnarray}
  \avav{\rho}_M &=& \int_{\sqrt{2gz}}^\infty \rho(z;v_z) P_M (v_z) \frac{t_D}{\av{t_D}}dv_z\\
  &=& \frac{1}{2H} e^{-z/2H} K_0\left({z \over 2H}\right)
  \label{avavrho_M}
\end{eqnarray}
where $K_0$ is the modified Bessel function of the second kind.
The column integrals are
\begin{eqnarray}
  \int_0^\infty \avav\rho_M \, dz &=& 1\\
  \int_0^\infty z \avav\rho_M  \, dz &=& \frac{2}{3} H  \label{avavz}
\end{eqnarray}
The median is determined numerically as $z_m \approx 0.39 H$.

For $z\ll H$, $K_0(z/2) = -\ln(z/4) - \gamma$, where $\gamma$ is the Euler constant, and in this limit
\begin{equation}
\avav\rho_M = -\frac{1}{2H} \left( \ln\left({z\over 4H}\right) + \gamma \right)  \quad{\rm for}\quad z \ll H
\end{equation}
which implies that the density near the surface goes to infinity at a logarithmic rate.
In the opposite limit of large height, $z\gg H$, $K_0(z/2) = \sqrt{\pi/z}\, e^{-z/2}$, and therefore
\begin{equation}
\avav\rho_M = \frac{1}{2} \sqrt{\pi\over z}\, e^{-z/H} \quad{\rm for}\quad z \gg H
\end{equation}
At large heights, the density falls off faster than an exponential.

\begin{figure*}[tb!]
  \centerline{\includegraphics[width=3.5in]{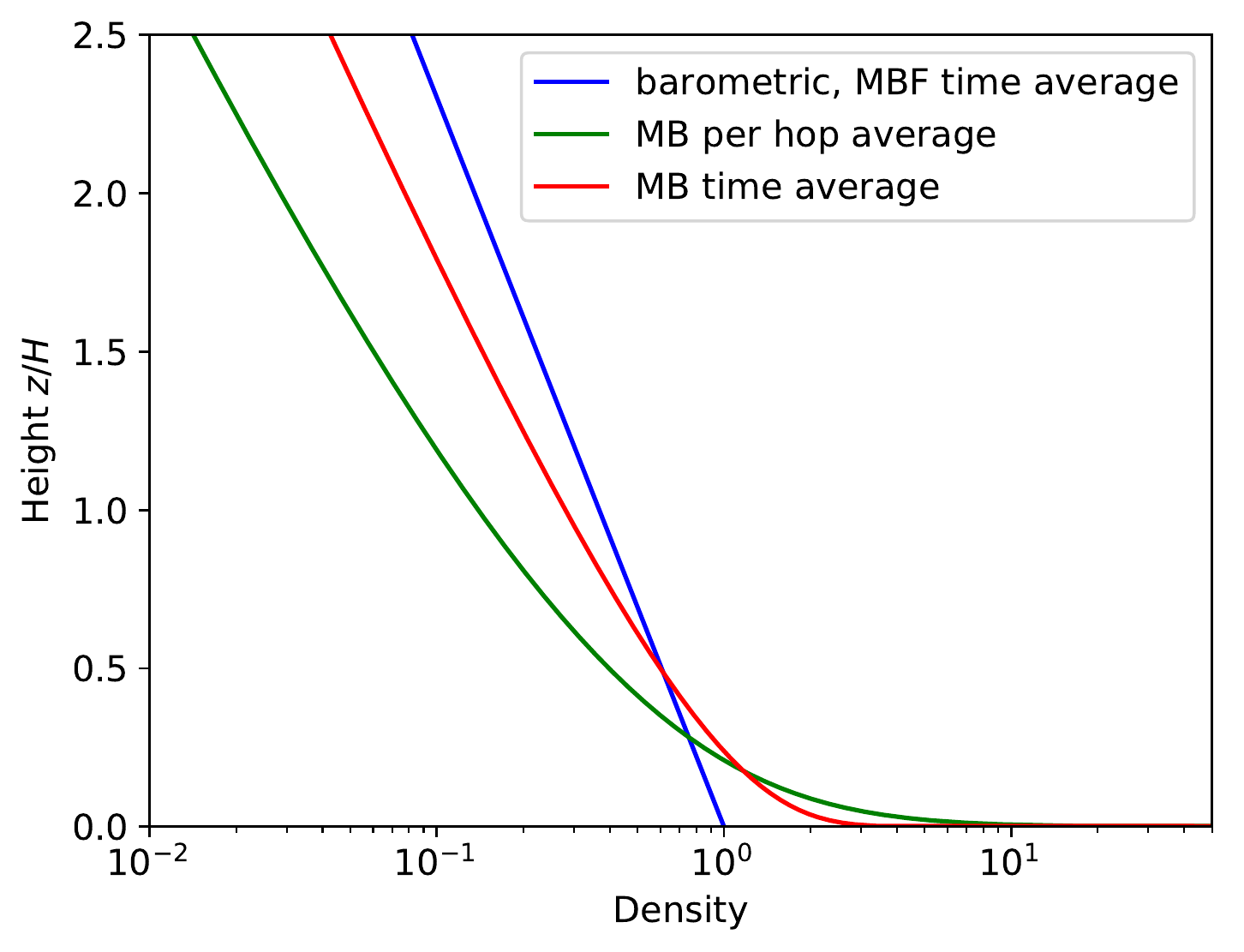}}
  \caption{Theoretical density profiles of exospheres according to eqs.\ (\ref{baro}), (\ref{avrho_M}), and (\ref{avavrho_M}).}
  \label{fig:densityprofiles}
\end{figure*}

Figure~\ref{fig:densityprofiles} compares $\av\rho_M$ and $\avav\rho_M$ with the barometric law.  What appears to be a ``ground-hugging'' population is actually part of a population described by a single temperature.

As apparent from eqs.\ (\ref{avrho0}) and (\ref{avavrho0}), the near-surface divergence arises from particles with small vertical launch velocities. Typically these will have small launch angles.
On a rough surface launches at small angles are blocked more often than launches at steep angles \cite{butler97}, which eliminates the divergence, but it does not necessarily imply the ground-hugging population disappears.

\subsection{Armand/MBF distribution: the classical density profile}

The Armand distribution has a basis in desorption chemistry \cite{armand77,hodges16}.
In its 3-dimensional form it is \cite{armand77} 
\begin{equation}
P_{3A} = 
  \left( \sqrt{\beta\over\pi} e^{-\beta v_x^2} \right)
  \left( \sqrt{\beta\over\pi} e^{-\beta v_y^2} \right)
  \left( 2\beta v_z e^{-\beta v_z^2} \right) 
  \label{A_cart}
\end{equation}
with $\beta=m/(2kT)$.
It is also known as ``Maxwell-Boltzmann flux'' (MBF) distribution \cite{smith78,hodges80b}, and chosen because the flux from an exobase involves an extra factor of $v_z$, due to projection of the velocities onto the vertical.
The distribution of the vertical component is given by
\begin{equation}
  P_A(v_z) = 2 \beta v_z e^{-\beta v_z^2}
  \label{Parmand}
\end{equation}
which has averages
\begin{equation}
  \av{v_z}_A = \frac{1}{2} \sqrt{\pi\over \beta}, \quad
  \av{v_z^2}_A = \frac{1}{\beta} = \frac{2kT}{m}, \quad
  \avav{v_z^2}_A = \frac32 \frac{1}{\beta}
\end{equation}
The MBF distribution implicitly allocates $kT$ for the vertical translational mode, instead of $kT/2$.
The lateral velocity components follow Boltzmann distributions.

The average of $t_D$, using eqs.\ (\ref{tD}), (\ref{av}), and (\ref{Parmand}), is
\begin{equation}
\av{t_D}_A = \frac{1}{g} \sqrt{\pi\over \beta } = \sqrt{2 \pi H \over g}
\end{equation}
The particle-averaged density profile is
\begin{eqnarray}
  \av{\rho}_A &=& \int_{\sqrt{2gz}}^\infty \rho(z;v_z) P_A (v_z) \, dv_z  \\
  &=& \frac{1}{2H} e^{-z/2H} K_0 \left(\frac{z}{2H}\right)
  \label{avrho_A}
\end{eqnarray}
The time average is
\begin{equation}
  \avav{\rho}_A = \int_{\sqrt{2gz}}^\infty \rho(z;v_z) P_A (v_z) \frac{t_D}{\av{t_D}_A}dv_z
  = \frac{1}{H} e^{-z/H}
  \label{avavrho_A}
\end{equation}
This reproduces the barometric formula (\ref{baro}), even with the same scale height $H$.

The classical particle-based \cite{opik59,shen63} and Liouville theorem based \cite{aamodt62} derivations lead to the same density profile and are equivalent to each other.
Here this derivation is reproduced for the simple case of constant gravitational acceleration.
Shen \cite{shen63} writes the density profile as
\begin{equation}
  \rho(z) = 4\pi \left({\beta\over\pi}\right)^{3/2} \!\int\!\!\!\int \frac{1}{\sqrt{v_z^2-2gz}} e^{-\beta (v_z^2+v_t^2)} v_t v_z dv_t dv_z
  \label{shenrho}
\end{equation}
This integrates to
\begin{eqnarray}
  \rho &=& 4\pi \left({\beta\over\pi}\right)^{3/2} \!\!
  \int_{\sqrt{2gz}}^\infty \!\!\!\! \frac{e^{-\beta v_z^2}}{\sqrt{v_z^2-2gz}}  v_z dv_z \int_0^\infty e^{-\beta v_t^2} v_t dv_t \nonumber \\
 & =& e^{-2\beta gz} = e^{-z/H}
\end{eqnarray}
This reproduces the barometric behavior (\ref{baro}) and $\avav{\rho}_A$, eq.\ (\ref{avavrho_A}), with different normalizations.

The density profile is an ensemble average over the inverse vertical velocity (\ref{halfform}).
Equation (\ref{shenrho}) is the ensemble average with respect to the MBF distribution.
The probability distribution (\ref{A_cart}) can be cast in the following form:
\begin{eqnarray}
 P_{3A}  dv_x dv_y dv_z 
 \!\!\! &=& \!\!
              {2\beta^2\over\pi} v_z e^{-\beta (v_t^2+v_z^2)} (v_t d\phi dv_t) dv_z  \nonumber \\
 \!\!\! &=& \!\!
              4 \beta^2 v_z e^{-\beta (v_t^2+v_z^2)} (v_t dv_t) dv_z \label{likeshen}
\end{eqnarray}
Equation (\ref{likeshen}) is a two-dimensional form of the MBF distribution and appears as integrand in (\ref{shenrho}).

\subsection{Armand/MBF for rough surface}

A planetary surface is rough on small scales, so the local surface normal rarely points along the direction of gravity. Roughness acts to make the velocity more isotropic.
The density profile can be calculated for emission from a vertical wall instead of from a horizontal surface.
For a vertical wall, it is the $y$-component that has a velocity prefactor. Instead of (\ref{A_cart}),
\begin{equation}
  P_{3A\perp} = 
  \left( \sqrt{\beta\over\pi} e^{-\beta v_x^2} \right)
  \left( 2\beta v_y e^{-\beta v_y^2} \right) 
   \left( 2 \sqrt{\beta\over\pi} e^{-\beta v_z^2} \right)
\end{equation}
where subscript $A\perp$ refers to the Armand distribution for a vertical wall.
The probability distribution of the vertical velocity component is
\begin{eqnarray}
  P(v_z)_{A\perp} &=&  \int_{-\infty}^\infty dv_x \int_0^\infty dv_y P_{3A} \nonumber \\
  &=& 2 \sqrt{\beta\over\pi} e^{-\beta v_z^2}
  = P(v_z)_M
\end{eqnarray}
which is a Boltzmann distribution, as in (\ref{PM1}).
Hence, the solution is the same as for the Maxwell-Boltzmann distribution, the vertical profile is (\ref{avavrho_M}), which diverges and has a ground-hugging behavior.

For a general slope angle, there does not appear to be a closed-form solution for the density profile, but the solution has the exponential density profile (\ref{avavrho_A}) and the faster-than-exponential profile (\ref{avavrho_M}) as end-members, respectively, for horizontal and vertical surfaces. For a rough surface, that contains both horizontal and vertical elements, the combined solution will have an exponential behavior for high altitudes and a faster-than-exponential (a ground-hugging) behavior near the surface.
In this sense, the exosphere can be thought of as consisting of two components: One that consists predominantly of fast particles ejected upwards from horizontal surfaces, and one that consists predominately of slow particles ejected laterally (but upwards) from steep surfaces. However, they both have the same temperature and they are end-members of the same population rather than two distinct populations.

In conclusion, particles launched from a rough surface according to the Armand/MBF distribution are expected to follow the classical density profile at high altitudes, but to have a more ground-hugging behavior at low altitudes.

\section{Discussion}

\subsection{Knudsen cosine law}

What really is the initial velocity distribution for thermal desorption? And, specifially, what is its angular dependence?
An argument for a ``cosine-law" (an MBF distribution) has been made for an ideal gas at a wall \cite{knudsen1909b,comsa68}, which implies that the direction in which a molecule rebounds from a solid wall is independent of the direction it approaches the wall.
(This is the opposite of the velocity correlation expected for specular reflection on a smooth surface.)
The Knudsen cosine law can be understood in terms of the dispersing microgeometry of the wall \cite{feres04}, but the same decorrelation is expected from thermalization during contact. In any case, deviations from the cosine-law have been measured \cite{comsa68}.
The derivation of the Armand distribution \cite{armand77}, on the other hand, is based on the oscillations of the atoms in the crystal.
It is also known that the desorption rate depends on the order of desorption \cite{dejong90}.
In conclusion, the velocity distribution depends on the properties of the surface and the type of adsorption, but in the absence of more detailed information, the MBF/Armand distribution (which obeys the Knudsen cosine law) is the go~to assumption (and a better assumption than the MB distribution).

\subsection{The vertical flux}

The saturation vapor density is defined as the density where the flux from the gas phase to the condensed phase equals the flux from the condensed phase to the gas phase. In an exosphere, the mean free path is much larger than the scale height, so having a local density higher than the saturation vapor density is not unphysical. In other words, the states occupied by the H$_2$O molecules are not confined to near the surface, as they would be if the mean free path was much shorter.

For a water exosphere above an ice-covered surface, the flux on the surface is given by the sublimation rate of ice into vacuum.
In the stationary case, the net flux is zero, as an equal number of particles move up and down.
The uni-directional flux carries a prefactor of 1/2: $F=(1/2) \rho (dz/dt)$.
The ensemble-averaged flux 
decreases with height, because particles with different initial velocities reach different maximum heights: 
\begin{eqnarray} 
  \avav{F} &=& \frac12 \avav{\rho \frac{dz}{dt}} = \frac12 \avav{{g \over v_z}} \\
  &=& \frac{g}{2\av{v_z}} \int^\infty_{\sqrt{2gz}} P(v_z) \, dv_z 
\end{eqnarray}

For a Maxwellian distribution, 
\begin{equation}
\avav{F}_M = \sqrt{ \frac{\pi}{2} \frac{g}{H} } \, \mathrm{Erfc}\left( \sqrt{z/H} \right)
\end{equation}
and therefore
\begin{equation}
\avav{F}_M(z=0) = \sqrt{ \frac{\pi}{2} \frac{g}{H} } < \infty
\label{Fmax_0}
\end{equation}
This flux has units of inverse time and has to be multiplied with the column abundance $\sigma$.
For the Armand distribution and a horizontal surface
\begin{eqnarray}
\avav{F}_A &=& \sqrt{ \frac{2}{\pi} \frac{g}{H} } e^{-z/H} \label{Fmax_0_A} \\
&=&  \sqrt{2gH \over \pi} \, \avav\rho_A  = \av{v_z}_M \avav\rho_A
\end{eqnarray}
The Armand distribution has the special property that the flux is proportional to the density.

The sublimation rate of ice into vacuum is \cite{watson61b}
\begin{equation}
  E = { p_s \over \sqrt{2\pi k T m} }
  \label{Eice}
\end{equation}
where $p_s$ is the saturation vapor pressure.
For a gravitationally-bound water exosphere above an ice-covered surface
$E = \sigma \times \avav{F}_A(z=0)$.
Using (\ref{H}), (\ref{Fmax_0_A}), and (\ref{Eice})
\begin{equation}
\sigma = {p_s \over 2 m g}
\end{equation}
Hence, a simple relation is obtained for the column abundance of the water exosphere of an ice-covered body.

For example, for an ice-filled cold trap on the Moon at a temperature of 110~K,
$\sigma \approx 2\times 10^{13}$~molecules/m$^2$.
All cold traps on the Moon are smaller than the mean hop distance of water molecules \cite{schorghofer15}, which leads to dilution, and their interiors are often colder than 110~K, so this value of $\sigma$ is an upper estimate.

\subsection{Observed ``two-component'' exospheres}

Measurements of the vertical density profile of atomic H are available from the UV spectrometers on the Mariner 10 mission to Mercury. Shemansky and Broadfoot \cite{shemansky77rev}, and other authors, interpreted the observed profile as two populations of different temperature (420~K and 110~K), but there is no satisfactory physical explanation why there should be such a cold population at the subsolar point.
Vervack et al.\ \cite{vervack18abs} provided results for the H exosphere based on MESSENGER data, and also argue for a two-component system (100~K and 400~K).
For an Armand distribution on a rough surface or a non-Armand distribution on a flat surface, these density profiles may potentially be explained by a single population with a single temperature.
The numerical model results of Refs.\ \cite{wurz03,hurley16} reproduce the near-surface excess of H and He on Mercury.

The analysis here was focused on thermal desorption and gravitationally-bound exospheres. However, the lesson about the near-surface behavior of the density profile may well apply more broadly.

\section{Conclusions}

The vertical density profile of a stationary surface-bounded exosphere in a uniform gravity field was calculated for various probability distributions for the launch velocities. The two are connected through eq.~(\ref{newform}).

For a Boltzmann distribution, the profile is given by eq.~(\ref{avavrho_M}); the density approaches infinity near the surface and decays faster than exponentially at great height. 
For an Armand (Maxwell-Boltzmann-Flux) distribution, it follows an exponential.
More generally, if $P(v_z)$ is the probability distribution of the vertical component of launch velocities (and the function is analytic), then $P(0)>0$ implies the surface density diverges.
In other words, only probability distributions that inhibit relatively small vertical launch velocities do not create a ground-hugging population, eq.\ (\ref{avavrho0}).

For molecules desorbed from a vertical wall according to the Armand distribution, the density profiles also diverges near the surface.
The Armand distribution on a rough surface or any distribution with $P(v_z=0)>0$ on a horizontal surface produce a ground-hugging population.
These exact solutions may explain observed density profiles that have been interpreted as two-component exospheres in terms of a single component.

\newlength{\bibitemsep}\setlength{\bibitemsep}{.2\baselineskip plus .05\baselineskip minus .05\baselineskip}
\newlength{\bibparskip}\setlength{\bibparskip}{0pt}
\let\oldthebibliography\thebibliography
\renewcommand\thebibliography[1]{%
  \oldthebibliography{#1}%
  \setlength{\parskip}{\bibitemsep}%
  \setlength{\itemsep}{\bibparskip}%
}


\end{multicols}
\end{document}